\documentclass[twocolumn,showpacs,preprintnumbers,amsmath,amssymb]{revtex4}

\usepackage{graphicx}
\usepackage{dcolumn}
\usepackage{bm}
\usepackage{amssymb}
\usepackage{graphicx}
\usepackage{amsmath}
\usepackage{xspace}

\begin{document}

\title{Raman phonons in ReOFeAs (Re= Sm, La)}

\author{V. G. Hadjiev,$^1$ M. N. Iliev,$^2$ K. Sasmal,$^2$ Y. -Y. Sun,$^2$ and C. W. Chu$^{2,3,4}$}
\affiliation{$^1$Texas Center for Superconductivity and Department
of Mechanical Engineering, University of Houston, Texas
77204-5002, USA}
\affiliation{$^2$Texas Center for Superconductivity and Department
of Physics, University of Houston, Houston, Texas 77204-5002}
\affiliation{$^3$Hong Kong University of Science and Technology,
Clear Water Bay, Kowloon, Hong Kong}
\affiliation{$^4$Lawrence Berkeley National Laboratory, 1
Cyclotron Road, Berkeley, California 94720}

\date{April 14, 2008}

\begin{abstract}
We report the polarized Raman spectra of undoped ReOFeAs (Re = Sm,
La) collected at room temperature from $ab$ surfaces of impurity
free microcrystals. The spectra exhibit sharp phonon lines on very
weak electronic scattering background. The frequency and symmetry
of the four Raman phonons involving out-of-plane atomic vibrations
are found at 170~cm$^{-1}$ ($A_{1g}$, Sm), 201~cm$^{-1}$
($A_{1g}$, As), 208~cm$^{-1}$ ($B_{1g}$, Fe), 345~cm$^{-1}$
($B_{1g}$, O) for SmOFeAs, and 162~cm$^{-1}$ ($A_{1g}$, La),
208~cm$^{-1}$ ($A_{1g}$, As), 201~cm$^{-1}$ ($B_{1g}$, Fe),
316~cm$^{-1}$ ($B_{1g}$, O) for LaOFeAs.
\end{abstract}

\maketitle




Recent report of superconductivity at 26 K in LaO$_{1-x}$F$_x$As
(x=0.05 - 0.12)\cite{kamihara} has triggered an intense wave of
research activities comparable to that in the early days of the
superconducting cuprates and MgB$_2$. Soon after, other members of
the Fe-As oxypnictides family were found superconducting at even
higher temperatures.\cite{chen} The calculated electron-phonon
coupling\cite{singh,boeri} is found weak to produce a
superconducting state within the Eliashberg theory at the
experimentally measured $T_c$, and unconventional origin of
superconductivity mediated by antiferromagnetic spin fluctuations
is suggested.\cite{mazin} Although the phonons may play little
role in mediating the superconductivity in oxypnictides, their
study using Raman spectroscopy can provide important information
on the superconducting state through phonon coupling to the Raman
active electronic excitations.\cite{hadjiev, devereaux}

In this Communication we present the results of a polarized Raman
study of ReOFeAs (Re = Sm, La). The Raman spectra were obtained
under a microscope from very small plate-like single crystals
within polycrystalline samples, which allowed reliable
measurements only of the non-degenerated Raman modes in ReOFeAs.
The experimentally determined symmetry and frequencies of the
Raman active phonons are compared with those predicted by a
group-theoretical analysis of the $\Gamma$-point phonon modes and
recently reported {\it ab initio} calculations.\cite{singh,boeri}




LaOFeAs and SmOFeAs crystallize in the $P4/nmm$ (space group
No.129) structure.\cite{kamihara,chen} In Table I are given the
number of the expected $\Gamma$-phonons, their symmetry and the
corresponding Raman tensors.\cite{russo} The eigenvectors of the
$A_{1g}$ and $B_{1g}$ modes are displayed in Fig. 1. These
non-degenerated modes involve predominantly out-of-plane atomic
vibrations. The $E_g$ modes eigenvectors (not shown in Fig. 1) are
parallel to the $ab$-plane.

\begin{table}[htb]
\caption []{Wyckoff positions and irreducible representations
($\Gamma$-point phonon modes) for ReOFeAs (space group $P4/nmm$,
No.129, origin choice 2, $Z=2$). The Raman tensors are given in an
orthogonal system with $z$ and $x$ directions along the $C_4$ and
$C'_2$ axes, respectively.}
\begin{ruledtabular}
\begin{tabular}{c c c  }

    & Wickoff         & $\Gamma$-point \\
Atom    & position     & phonon modes   \\
\hline
  &  &    \\
Sm/La  & 2c        & $A_{1g} + A_{2u} + E_g + E_u$  \\
O  & 2a       & $B_{1g} + A_{2u} + E_g + E_u$   \\
Fe  & 2b         & $B_{1g} + A_{2u}+ E_g + E_u$  \\
As  & 2c       & $A_{1g} + A_{2u}+ E_g + E_u$ \\
  &  &    \\
\multicolumn{3}{c}{Modes classification:} \\

\multicolumn{3}{c}{$\Gamma_{\rm Raman} = 2A_{1g}+2B_{1g}+4E_g$}\\
\multicolumn{3}{c}{$\Gamma_{\rm IR} = 3A_{2u}+3E_u$}\\
\multicolumn{3}{c}{$\Gamma_{\rm Acoustic} = A_{2u} + E_u$}\\
  &  &  \\
\multicolumn{3}{c}{Raman tensors:} \\

\multicolumn{3}{c}{$A_{1g}(x^2+y^2,z^2) \rightarrow \left[
\begin{array}{ccc}
 a & 0 & 0 \\
 0 & a & 0 \\
 0 & 0 & b
 \end{array}  \right]$} \\

 & \\

 \multicolumn{3}{c}{$B_{1g}(x^2-y^2) \rightarrow \left[
\begin{array}{ccc}
 c & 0 & 0 \\
 0 & -c & 0 \\
 0 & 0 & 0
 \end{array}  \right]$}\\

  & \\

 \multicolumn{3}{c}
 {$E_{g_1}(xz),E_{g_2}(yz) \rightarrow \left[
\begin{array}{ccc}
 0 & 0 & -e \\
 0 & 0 & 0 \\
 -e & 0 & 0
 \end{array}\right]$,
 $\left[ \begin{array}{ccc}
 0 & 0 & 0 \\
 0 & 0 & e \\
 0 & e & 0
 \end{array}\right]$ }  \\
    &  \\

\end{tabular}
\end{ruledtabular}
\end{table}




The polycrystalline pellets of LaOFeAs and SmOFeAs used in our
experiment were prepared by solid state sintering. Firstly, La
(Sm) metal powders and As chips were mixed and pressed into
pellet, sealed in vacuumed quartz tube, and heated at $1000^{\rm
o}$ C for 20-50 hrs. The formed LaAs (SmAs) was then mixed with Fe
powder and ${\rm Fe_2O_3}$ according to the designed
stoichiometry, pressed into pellets and sealed in quartz tubes
again. The final reaction was carried out at $1150^{\rm o}$ C for
60 hrs. The powder X-ray diffraction of the LaOFeAs phase revealed
a tetragonal structure with room temperature lattice constants of
$a = 0.4021$ nm and $c = 0.8723$ nm. The lattice constants of the
SmOFeAs phase were $a = 0.3942$ nm and $c = 0.8498$ nm.

The Raman spectra of small single crystals were measured under a
microscope ($\times 100$-magnification) attached to a Horiba JY
T64000 triple spectrometer. Both compounds, SmOFeAs and LaOFeAs,
appeared to be highly absorbing in the visible range and having
relatively poor thermal conductivity that required use of incident
laser power density below $10^4$~W/cm$^2$. In addition, phonon
scattering intensity was found to be relatively weak,  about
~$10^{-3}$ of that of Si. The Raman spectra presented here were
excited with the 632.8~nm laser line. Raman measurements with the
514.5~nm excitation confirmed that the collected Raman spectra are
intrinsic.

\begin{figure}[htb]
\includegraphics[width=7cm]{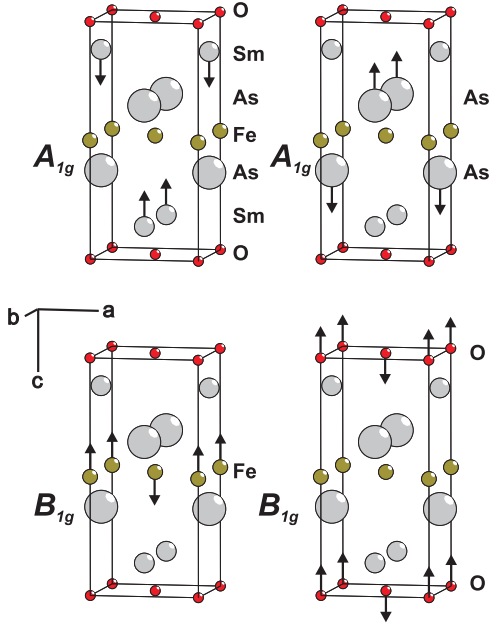}\\

\caption{(Color online) Eigenvectors of all non-degenerated Raman
phonons of ReOFeAs (Re = Sm, La).}
\end{figure}

The SmOFeAs and LaOFeAs microcrystals have plate-like form, which
by analogy with other layered compounds suggests that the larger
faces most likely are parallel to the crystallographic $ab$-plane,
(001). Under this assumption and considering the Raman tensors
$\mathbf{R}$ given in Table I, one should expect the Raman mode
intensity $I(\varphi) \propto [\vec{e}_s \mathbf{R} \vec{e}_i]^2$
($\vec{e}_i$ and $\vec{e}_s$ are the incident and scattered light
polarizations) to vary with the angle $\varphi$ between the [100]
crystallographic direction (along the $a$-axis) and the incident
light polarization as $ I^\parallel_{A_{1g}}(\varphi) \propto a^2
$ and $I^\perp_{A_{1g}}(\varphi) =0$ for parallel $\vec{e}_i
\parallel \vec{e}_s$ and crossed $\vec{e}_i \perp \vec{e}_s$ scattering
configurations, respectively. For the $B_{1g}$ modes the expected
intensity dependencies are $I^\parallel_{B_{1g}}(\varphi) \propto
c^2\cos^22\varphi$ and $I^\perp_{B_{1g}}(\varphi) \propto
c^2\sin^2\varphi\cos^2\varphi$. No $E_g$ modes should be
observable with incident and scattered light polarizations in the
$ab$-plane.

\begin{figure}[htb]
\includegraphics[width=7cm]{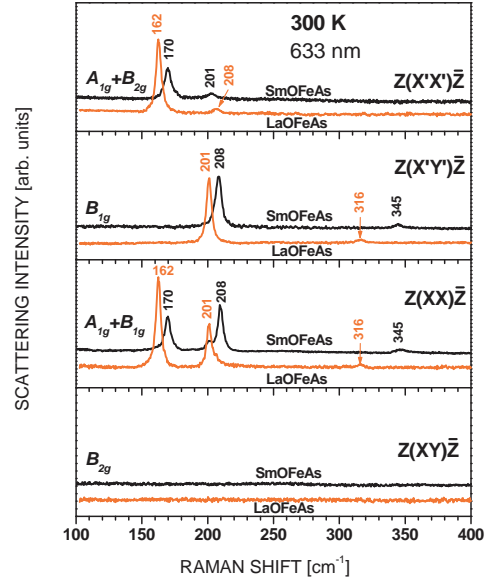}\\

\caption{(Color online) Polarized Raman spectra of SmOFeAs and
LaOFeAs measured in back-scattering configuration with incident
and scattered light polarizations in the $ab$-plane. Notations:
$Z\parallel [001]$, ${\bar Z}\parallel [00{\bar 1}]$, $X\parallel
[100]$, $Y\parallel [001]$, $X'\parallel [110]$, and $Y'\parallel
[{\bar 1}10]$.}
\end{figure}

Figure 2 compares the polarized Raman spectra of SmOFeAs and
LaOFeAs measured at room temperature in back scattering
configurations with light polarizations along certain
crystallographic directions. The scattering configurations are
presented in Porto notation.\cite{notation} The $Z$ direction is
along [001] ($c$-axis), $X$ along [100] ($a$-axis), and $X'$ is
parallel to [110]. The $Y$ and $Y'$ directions are orthogonal to
$X$ and $X'$. The Raman spectra in Fig. 2 were taken from square
shaped plates (10~x~10~$\mu{\rm m}^2$), which were found to obey
thoroughly the angular dependencies
$I^\parallel_{A_{1g}/B_{1g}}(\varphi)$ and
$I^\perp_{A_{1g}/B_{1g}}(\varphi)$ for light polarizations in the
$ab$-plane. The assignment of the phonon lines in Fig. 2 is
straightforward. The measured non-degenerated modes in SmOFeAs
are: 170~cm$^{-1}$ ($A_{1g}$, Sm), 201~cm$^{-1}$ ($A_{1g}$, As),
208~cm$^{-1}$ ($B_{1g}$, Fe), and 345~cm$^{-1}$ ($B_{1g}$, O). For
LaOFeAs the corresponding modes are 162~cm$^{-1}$ ($A_{1g}$, La),
208~cm$^{-1}$ ($A_{1g}$, As), 201~cm$^{-1}$ ($B_{1g}$, Fe),
316~cm$^{-1}$ ($B_{1g}$, O).

The phonon dispersions and electron-phonon coupling of LaOFeAs
have already been reported as calculated using the Quantum
Espresso code with ultrasoft pseudopotentials.\cite{singh,boeri}
The calculations predict phonon branches with little dispersion in
the z direction ($c$-axis), reflecting the layered structure of
LaOFeAs, and phonon spectrum that spreads up to 500~cm$^{-1}$. The
oxygen vibrations are expected between 300~cm$^{-1}$ and
500~cm$^{-1}$, whereas those of La, Fe, and As are occupying the
range below 300~cm$^{-1}$.  No assignment of the $\Gamma$-point
phonons, however, is made in Refs.~\cite{singh,boeri}. Given the
measured non-degenerated Raman phonon frequencies in LaOFeAs are
between 150~cm$^{-1}$ and 350~cm$^{-1}$ we can assign them to the
four non-degenerated dispersion curves crossing the $\Gamma$-point
near 180~cm$^{-1}$, 200~cm$^{-1}$, 217~cm$^{-1}$, and
311~cm$^{-1}$ in the calculated phonon dispersion curves.
Therefore, the calculated frequencies are in fairly good agreement
(less then 10\% deviation) with the experimental ones.

Further, we briefly compare the Raman phonons of SmOFeAs with
those of LaOFeAs. The only atomic substitution in these
isostructural compounds is at the La site. Sm ($m_{\rm Sm}=
150.4$~u) is heavier than La ($m_{\rm La}=139$~u), however, the
smaller ionic radius of Sm$^{3+}$ results in shortening of the
Sm-O distances ($d_{\rm Sm-O}=2.259\,\AA $) by 7\% compared to
those of La-O ($d_{\rm La-O}=2.359\,\AA $) in
LaOFeAs.\cite{bondlength} Using the standard dependence of the
phonon frequencies on the atomic masses and nearest neighbor
distances in isostructural compounds,\cite{atanasova} we can write
$\omega_{\rm Sm}/\omega_{\rm La}\approx(m_{\rm La}/m_{\rm
Sm})^{1/2}({d_{\rm La-O}/d_{\rm Sm-O}})^{3/2}$. From this
expression we find that $\omega_{\rm Sm}\approx1.026\,\omega_{\rm
La}$, which is close to the experimentally established
$\omega_{\rm Sm}\approx1.049\,\omega_{\rm La}$. The shorter Sm-O
distance also causes a higher $B_{1g}$ oxygen mode frequency in
SmOFeAs. Note that Fe and As mode frequencies are close in both
compounds but while $\omega_{As}>\omega_{Fe}$ in LaOFeAs, this
relation is opposite for SmOFeAs. We attribute this behavior of
the phonon modes to the particular eigenvector of the As atoms
vibrating between Fe and La/Sm sublattices (see Fig. 1), whereas
Fe is vibrating only within the As sublattice. It seems that
although the Fe-As distance decreases in going from LaOFeAs to
SmOFeAs, the softening of the As mode is due to the compensatory
increase in the Sm-As distance.

The Raman detection of the $E_g$ modes was challenging because the
laser spot was comparable in size with the dimension of the
crystal $ac$-surface and no reliable results were produced.




In conclusion, we measured and assigned the four non-degenerated
Raman phonons in undoped ReOFeAs (Re = Sm, La). The phonon Raman
lines are very sharp indicating small if any renormalization due
to interactions with the other excitations in these compounds.

\acknowledgments This work was supported in part by the State of
Texas through the Texas Center for Superconductivity at the
University of Houston (TcSUH), the T. L. L. Temple Foundation, the
John J. and Rebecca Moores Endowment, and the United States Air
Force Office of Scientific Research.


\begin{thebibliography}{99}
\bibitem{bondlength} The distances were calculated for crystal
structures with following parameters: $a=0.4021$~nm,
$c=0.8723$~nm, $z_{\rm La}=0.1415$, and $z_{\rm As}=0.6512$ for
LaOFeAs; $a=0.3942$~nm, $c=0.8498$~nm, $z_{\rm Sm}=0.1299$, and
$z_{\rm As}=0.6402$ for SmOFeAs.
\end{thebibliography}
\end{document}